\documentclass[floatfix,showpacs,prl,twocolumn]{revtex4}
\usepackage{amsbsy,amssymb,amsmath,bm}
\usepackage{graphicx,color}
\usepackage{amsfonts}
\usepackage{amssymb}
\usepackage{amsmath}

\newcommand{\U}[1]{\sigma}

\begin{document}

\title{Chiral anomaly and strength of the electron-electron interaction in
graphene}
\author{B. Rosenstein$^{1,2}$,M. Lewkowicz$^{2}$,T. Maniv$^{3}$}
\affiliation{\textit{$^{1}$Electrophysics Department, National Chiao Tung University,
Hsinchu 30050,} \textit{Taiwan, R. O. C}.}
\affiliation{\textit{$^{2}$Physics Department, Ariel University Center of Samaria, Ariel
40700, Israel}}
\affiliation{$^{2}$Schulich Faculty of \textit{Chemistry, Technion, Haifa 32000, Israel}}
\email{maniv@tx.technion.ac.il}
\date{\today }

\begin{abstract}
The long standing controversy concerning the effect of electron - electron
interaction on the electrical conductivity of an ideal graphene sheet is
settled. Performing the calculation directly in the tight binding approach
without the usual prior reduction to the massless Dirac (Weyl) theory, it is
found that, to leading order in the interaction strength $\alpha
=e^{2}/\hbar v_{0}$, the DC conductivity $\sigma /\sigma _{0}=1+C\alpha
+O\left( \alpha ^{2}\right) $ is significantly enhanced with respect to the
independent-electrons result $\sigma _{0}$, i.e. with the value $C=0.26$. \
The ambiguity characterizing the various existing approaches is nontrivial
and related to the chiral anomaly in the system. In order to separate the
energy scales in a model with massless fermions, contributions from regions
of the Brillouin zone away from the Dirac points have to be accounted for.
Experimental consequences of the relatively strong interaction effect are
briefly discussed.
\end{abstract}

\maketitle

\textit{Introduction. }It has been demonstrated recently that a graphene
sheet, especially one suspended on leads, is one of the purest electronic
systems. The scattering of charge carriers in suspended graphene samples of
submicron length is so negligible that the transport is ballistic \cite%
{Andrei,Geim11}. The novelty of the physics of undoped graphene is in the
ability to probe the "ultrarelativistic" physics of excitations leading to
numerous similarities with phenomena previously associated with the high
energy physics. Examples include Zitterbewegung and Klein tunneling\cite%
{Katsnelson}, electron - hole (Schwinger) pair creation by an electric field 
\cite{Lewkowicz}, a possibility of dynamical (chiral) symmetry breaking by
electron interaction effects\cite{Drut} (exciton condensation) and the
chiral (parity) anomaly\cite{Semenoff}. The latter, a quantum anomaly,
attributed to graphene long before its discovery, is one of the most
remarkable features of a relativistic field theory with massless fermions%
\cite{Brown,anomaly}. Generally it is associated with the fact that a
classical symmetry is "broken" by quantization in the case of an infinite
number of degrees of freedom, when the ultraviolet (UV) cutoff is necessary.
Chiral anomaly means that the classical axial $U\left( 1\right) $ symmetry
is violated. This led to explanations of such physical phenomena\cite{Brown}
as $\pi ^{0}\rightarrow \gamma \gamma $ decay (that would be suppressed by
the symmetry), the solution of the problem of the large mass of the $\eta $
meson (zero, if it were to be a Goldstone boson of a nonanomalous symmetry)
etc. The anomalies are notorious in that calculations of a well defined
physical quantity using different UV cutoff procedures (for example the
sharp momentum cutoff, lattice regularization or a properly defined
dimensional regularization) led to different finite values. The physical
essence of this ambiguity is that there is no simple separation between the
UV and infrared (IR) physics and certain care should be exercised in
construction of the correct effective low energy model. This might be
suspected to occur in theory of graphene. In description of graphene, while
the starting point might be an atomic or tight binding model\cite{Castro},
one typically replaces it by an massless effective Dirac (Weyl) model "near"
its two Dirac points constituting the Fermi "surface" of undoped graphene.

In this note we point out that the elucidation of the ambiguities
encountered in the theory of the (\textit{apriori} strong) Coulomb
interactions should be associated with a careful treatment of the separation
of scales due to the anomaly. We show in detail, using the tight binding
model providing a natural UV cutoff, that some aspects of the graphene
physics are \textit{not} dominated by the two Dirac points of the Brillouin
zone at which the spectrum is gapless. \ The low frequency conductivity in
the quasi-dielectric phase below the exciton condensation critical coupling%
\cite{Drut,Vozmediano} $\alpha \equiv e^{2}/\hbar v<\alpha _{c}$ (neglecting
weak logarithmic renormalization of the electron velocity \cite%
{Vozmediano,Gonzalez,Mishchenko,remark}, $v=v_{0}\sim 10^{6}m/s)$, is given
in terms of its value in the noninteracting theory, $\sigma
_{0}=e^{2}/4\hbar $, by

\begin{equation}
\sigma \left( \omega \right) /\sigma _{0}=1+C\alpha +O\left( \alpha
^{2}\right) \text{.}  \label{sigma}
\end{equation}%
This expression is valid for frequencies below the hopping energy $\gamma
=2.7eV$. The static dielectric constant is therefore given by $\varepsilon
^{-1}=1-\pi \alpha /2\left( 1+C\alpha \right) +O\left( \alpha ^{3}\right) $.
The value of the only numerical constant $C$ appearing here has been a
matter of intense controversy. The first detailed calculation by Herbut,
Juricic and Vafek\cite{Herbut1} utilizing a sharp momentum cutoff
regularization of the Dirac model provided a value of order $1$: 
\begin{equation}
C^{\left( 1\right) }=\frac{25}{12}-\frac{\pi }{2}\approx 0.51\text{.}
\label{c1}
\end{equation}%
The use of the sharp momentum cutoff was criticized by Mishchenko\cite%
{Mishchenko2}, who obtained a value of 
\begin{equation}
C^{\left( 2\right) }=\frac{19}{12}-\frac{\pi }{2}\approx 0.01  \label{c2}
\end{equation}%
making a "soft" momentum cutoff regularization. He supported this choice by
the consistency of the Kubo and the kinetic equation calculations of
conductivity with that of the polarization function (dielectric constant).
The consistency required a modification of the long range interaction so
that it becomes UV cutoff dependent. It was further supported by Sheehy and
Schmalian\cite{Schmalian2} who used yet a different cutoff procedure and
pointed out that only such a small value of $C$ can explain the experimental
observation of the optical conductivity in graphene on a substrate\cite%
{GeimScience08}, which is\ within 1\% of $\sigma _{0}$. This apparently
closed the issue. Albeit such a small numerical value would have profound
physical consequences even beyond the transport and dielectric properties.

Nevertheless the interaction strength $C$ was recalculated once again by
Vafek, Juricic and Herbut\cite{Herbut2} who argued that the modification of
the interaction requires simultaneously a Pauli - Villars regularization of
massless fermions. They applied yet another regularization, making the space
dimensionality fractional, $D=2-\varepsilon $ (similar to the space - time $%
4-\varepsilon $ regularizations that has been long in use in high energy and
critical phenomena physics\cite{anomaly}) that modified both the current
operator and the interaction in such a way that they satisfy the Ward
identities and obtained%
\begin{equation}
C^{\left( 3\right) }=\frac{11}{6}-\frac{\pi }{2}\approx 0.26\text{.}
\label{c3}
\end{equation}%
The dimensional regularization is questionable on physical grounds and in a
comprehensive subsequent work\cite{MacDonald} the authors reaffirmed the
small value $C^{\left( 2\right) }$ and it seems that it is a commonly
accepted one. To refute the earlier calculation of ref.\cite{Herbut2} they
write "of course, satisfying the Ward-Takahashi identity does not guarantee
that the regularization scheme will produce the exact value of $C$ for the
physical system. We believe that if a really quantitative result is desired
for the constant $C$, then one should resort to a complete electronic
structure calculation (based, for example, on a realistic tight-binding
Hamiltonian) rather than working with an effective low-energy theory". We
followed this path, but surprisingly found that \textit{the tight binding
value is} $C^{\left( 3\right) }$. The situation is further complicated by
other values in literature like $C=0.34$ obtained in a dielectric constant
calculation\cite{Castro2}.

To reveal the origin of the ambiguity exhibited by the various values of $C$
(there is a consensus that all the calculations are mathematically sound\cite%
{Schmalian2}), we use a dynamical approach (used previously to address the
ambiguity of the noninteracting case\cite{Lewkowicz,Ziegler}) directly in
the DC case by "switching on" a uniform electric field in the tight binding
model with Coulomb interactions, and then considering the large-time limit.
This approach (known in field theory as the "infinite hotel story") is the
best way to reveal physical effects of anomalies\cite{Brown,anomaly}. One
can directly separate the contributions from the neighborhood of Dirac
points and the "anomalous" contributions from the rest of the Brillouin
zone, so that one can decide what regularization of the effective Weyl
theory is the correct one. We have also performed a standard diagrammatic
Kubo formula calculation of the general AC conductivity within the tight
binding model and obtained the same result.

\textit{The tight binding model and its linear response to an electric
field. }Electrons in graphene are described sufficiently accurately for our
purposes by the 2D tight binding model of nearest neighbour interactions in
external field described by Wilson links\cite{Smit}: 
\begin{eqnarray}
K\left[ \mathbf{A}\right] &=&-\gamma \sum\limits_{\mathbf{n,}i}c_{\mathbf{r}%
_{\mathbf{n}}}^{\sigma \dagger }W_{\mathbf{n},i}c_{\mathbf{r}_{\mathbf{n}}+%
\mathbf{d}_{i}}^{\sigma }+hc;  \label{K} \\
W_{\mathbf{n},i} &=&\exp \left[ \frac{ie}{c\hbar }\int_{s=0}^{1}\mathbf{A}%
\left( \mathbf{r}_{\mathbf{n}}+s\mathbf{d}_{i},t\right) \cdot \mathbf{d}_{i}%
\right] \text{.}  \notag
\end{eqnarray}%
Here $\mathbf{A}\left( \mathbf{r},t\right) $ is the vector potential and $c_{%
\mathbf{r}}^{\sigma \dagger }$ creates an electron with spin $\sigma $
(summation over $\sigma $ implied) on the sites of the honeycomb lattice $%
\mathbf{r}_{\mathbf{n}}=n_{1}\mathbf{a}_{1}+n_{2}\mathbf{a}_{2}$, where
lattice vectors $\mathbf{a}_{1,2}$ and the nearest neighbours displacements $%
\mathbf{d}_{i}$ are defined in Fig.1.

\begin{figure}[tbp]
\begin{center}
\includegraphics[width=6cm]{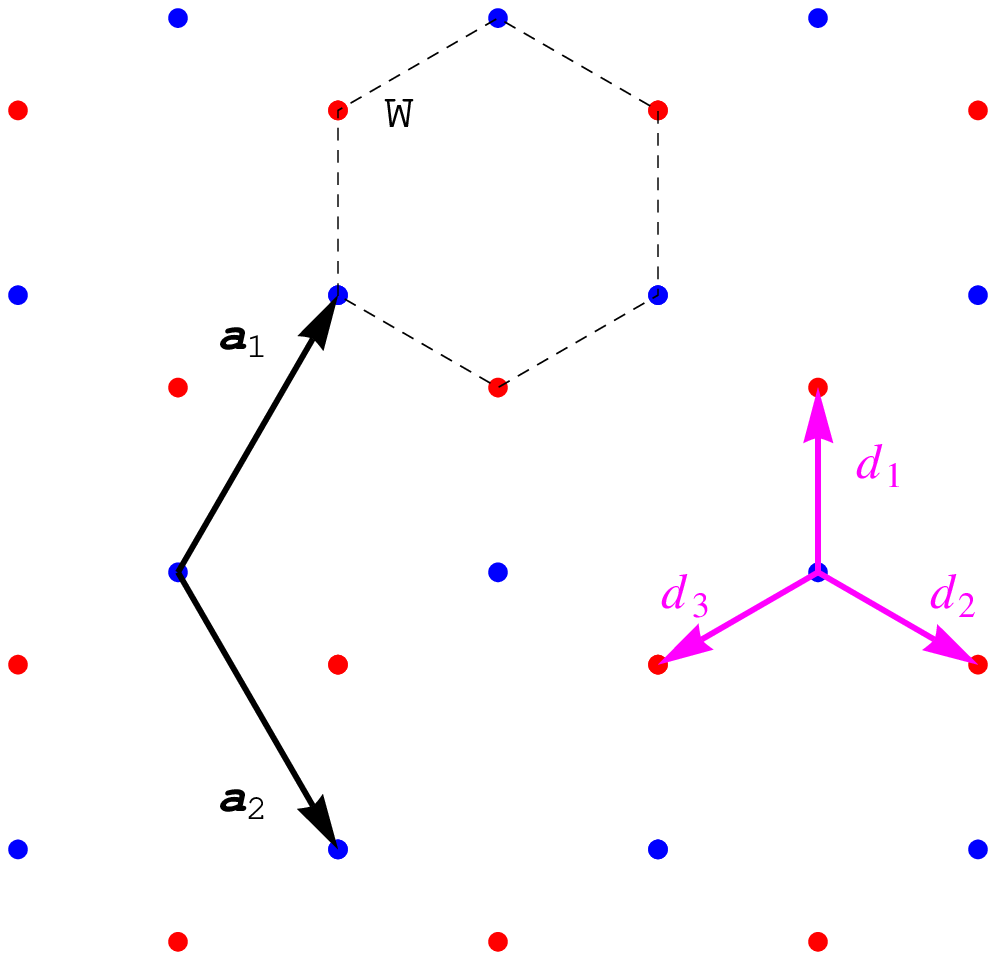}
\end{center}
\caption{Honeycomb lattice for graphene. The sublattice A (red) is spanned
by the lattice vectors $\mathbf{a}_{1,2}=$ $\frac{a}{2}\left( \pm 1,\protect%
\sqrt{3}\right) $ where $a\simeq 3\mathring{A}$. The three nearest
neighbours on sublattice B (blue) are displaced by $\mathbf{d}_{1}=\frac{1}{3%
}\left( \mathbf{a}_{1}-\mathbf{a}_{2}\right) $, $\mathbf{d}_{2}=\frac{1}{3}%
\left( \mathbf{a}_{1}+2\mathbf{a}_{2}\right) $,$\ \mathbf{d}_{3}=-\frac{1}{3}%
\left( 2\mathbf{a}_{1}+\mathbf{a}_{2}\right) $. Wilson links $W$ describing
the minimal coupling to the vector potential $\mathbf{A}\left( \mathbf{r}%
,t\right) $ are defined in Eq.(\protect\ref{K}).}
\end{figure}

Coulomb interactions between electrons are 
\begin{equation}
V=\sum\limits_{\mathbf{n,m}}\left[ 
\begin{array}{c}
\frac{1}{2}v\left( \mathbf{r}_{\mathbf{n}}-\mathbf{r}_{\mathbf{m}}\right)
\left( N_{\mathbf{n}}^{A}N_{\mathbf{n}}^{A}+N_{\mathbf{n}}^{B}N_{\mathbf{n}%
}^{B}\right) \\ 
+v\left( \mathbf{r}_{\mathbf{n}}-\mathbf{r}_{\mathbf{m}}-\mathbf{d}%
_{1}\right) N_{\mathbf{n}}^{A}N_{\mathbf{m}}^{B}%
\end{array}%
\right] \text{.}  \label{V}
\end{equation}%
where $N_{\mathbf{n}}^{A}=c_{\mathbf{r}_{\mathbf{n}}}^{\sigma \dagger }c_{%
\mathbf{r}_{\mathbf{n}}}^{\sigma }$\thinspace , $N_{\mathbf{n}}^{B}=c_{%
\mathbf{r}_{\mathbf{n}}+\mathbf{d}_{1}}^{\sigma \dagger }c_{\mathbf{r}_{%
\mathbf{n}}+\mathbf{d}_{1}}^{\sigma }$ and $v\left( \mathbf{r}\right)
=e^{2}/r$. The corresponding current density operator (in Heisenberg
picture) is $c\delta K\left[ \mathbf{A}\right] /\delta \mathbf{A}\left( 
\mathbf{r},t\right) $: 
\begin{eqnarray}
\mathbf{J}\left( \mathbf{r},t\right) &=&-\frac{i\gamma e}{\hbar }%
\sum\limits_{\mathbf{n},i}\mathbf{d}_{i}\int_{s=0}^{1}\delta \left( \mathbf{r%
}-\mathbf{r}_{\mathbf{n}}-s\mathbf{d}_{i}\right)  \label{J} \\
&&c_{\mathbf{r}_{\mathbf{n}}}^{\sigma \dagger }W_{\mathbf{n},i}c_{\mathbf{r}%
_{\mathbf{n}}+\mathbf{d}_{i}}^{\sigma }+hc\text{.}  \notag
\end{eqnarray}%
This describes a network-like flow of currents on links between neighboring
sites in Fig.1. As was emphasized in the context of quasi-local interaction
models (Hubbard models) in graphene in ref.\cite{Giuliani} (and much earlier
in the context of lattice gauge models of particle physics\cite{Smit}) this
model satisfies all the (nonanomalous) Ward identities associated with
charge conservation and therefore no trial-and-error modification of the
current operator is needed.

Let us consider a uniform electric field along the $y$ direction $\mathbf{E=}%
-\frac{d}{dt}\left( 0,A\left( t\right) /c\right) $ switched on at $t=0$. The
current density is expanded to first order in $\mathbf{A}$ as $\mathbf{J}=%
\mathbf{J}_{d}+\mathbf{J}_{p}$, with the relevant components being,

\begin{eqnarray}
J_{d}^{y}\left( \mathbf{r},t\right) &=&-\frac{e^{2}\gamma }{c\hbar ^{2}}%
A\left( t\right) \sum\limits_{\mathbf{n},i}\left( d_{i}^{y}\right) ^{2}c_{%
\mathbf{r}_{\mathbf{n}}}^{\sigma \dagger }c_{\mathbf{r}_{\mathbf{n}}+\mathbf{%
d}_{i}}^{\sigma }+hc,  \label{JpJd} \\
J_{p}^{y}\left( \mathbf{r},t\right) &=&\frac{ie\gamma }{\hbar }\sum\limits_{%
\mathbf{n},i}d_{i}^{y}c_{\mathbf{r}_{\mathbf{n}}}^{\sigma \dagger }c_{%
\mathbf{r}_{\mathbf{n}}+\delta _{\alpha }}^{\sigma }+hc\text{.}  \notag
\end{eqnarray}%
Averaging the expectation value of current density over the sample area $S$, 
$j\left( t\right) =\frac{1}{S}\int_{\mathbf{r}}\left\langle \phi \left\vert
J^{y}\left( t\right) \right\vert \phi \right\rangle $, one obtains:

\begin{eqnarray}
j_{d}\left( t\right) &=&-\frac{\gamma e^{2}}{c\hbar ^{2}S}A\left( t\right)
\sum\limits_{\mathbf{n,}i}\left( d_{i}^{y}\right) ^{2}Re\left\langle \phi
\left\vert c_{\mathbf{r}_{\mathbf{n}}}^{\sigma \dagger }c_{\mathbf{r}_{%
\mathbf{n}}+\mathbf{d}_{i}}^{\sigma }\right\vert \phi \right\rangle \mathbf{%
\mathbf{;}}  \label{jdjp} \\
j_{p}\left( t\right) &=&\frac{e^{2}}{\hbar S}\int_{t_{1}=0}^{t}A\left(
t_{1}\right) \sum\limits_{\mathbf{nm,}ij}d_{i}^{y}d_{j}^{y}  \notag \\
&&Im\left\langle \phi \left\vert c_{\mathbf{r}_{\mathbf{n}}}^{\sigma \dagger
}c_{\mathbf{r}_{\mathbf{n}}+\mathbf{d}_{i}}^{\sigma }e^{-iH\left(
t-t_{1}\right) }c_{\mathbf{r}_{\mathbf{m}}}^{\rho \dagger }c_{\mathbf{r}_{%
\mathbf{m}}+\mathbf{d}_{j}}^{\rho }\right\vert \phi \right\rangle \mathbf{%
\mathbf{\mathbf{.}}}  \notag
\end{eqnarray}%
The time independent Hamiltonian $H=K+V,$ $K\equiv K\left[ \mathbf{A}=0%
\right] $ and its ground state $\left\vert \phi \right\rangle $ are expanded
to first order in the interaction $V$. The tight binding model $K$ has a
spectrum $\varepsilon _{\mathbf{k}}=\pm \left\vert h_{\mathbf{k}}\right\vert 
$ determined by the structure function of the links $h_{\mathbf{k}}=-\gamma
\sum\nolimits_{i}$ $e^{-i\mathbf{k}\cdot \mathbf{d}_{i}}$. The DC field is
defined by $A\left( t\right) =-cEt$ and resuls of direct calculation are
presented and discussed in what follows.

\textit{The evolution of the current. }The current density to first order in
interactions is

\begin{equation}
j\left( t\right) =\sigma _{0}E\left[ C^{0}\left( t\right) +\alpha C\left(
t\right) \right]  \label{j(t)}
\end{equation}%
The components of the minimal dimensionless conductivity $C^{0}\left(
t\right) $ are written as integrals over the Brillouin zone (BZ), see Fig.2:

\begin{eqnarray}
C_{d}^{0}\left( t\right) &=&-16\hbar t\sum\nolimits_{\mathbf{k}}Re\left( 
\frac{h_{\mathbf{k}}^{\ast }h_{\mathbf{k}}^{\prime \prime }}{\varepsilon _{%
\mathbf{k}}}\right) \mathbf{\mathbf{;}}  \label{s0} \\
C_{p}^{0}\left( t\right) &=&-16\hbar t\sum\nolimits_{\mathbf{k}}\frac{\zeta
_{\mathbf{k}}^{2}}{\varepsilon _{\mathbf{k}}}-8\hbar \sum\nolimits_{\mathbf{k%
}}\frac{\zeta _{\mathbf{k}}^{2}}{\varepsilon _{\mathbf{k}}^{2}}\sin \left(
2\varepsilon _{\mathbf{k}}t/\hbar \right) \text{.}  \notag
\end{eqnarray}%
where $\zeta _{\mathbf{k}}=Im\left( z_{\mathbf{k}}h_{\mathbf{k}}^{\prime
}\right) \mathbf{\mathbf{\mathbf{,}}}$ $z_{\mathbf{k}}=h_{\mathbf{k}}^{\ast
}/\varepsilon _{\mathbf{k}},$ and a prime denotes a derivative with respect
to momentum along the field, $k_{y}$.

\begin{figure}[tbp]
\begin{center}
\includegraphics[width=8cm]{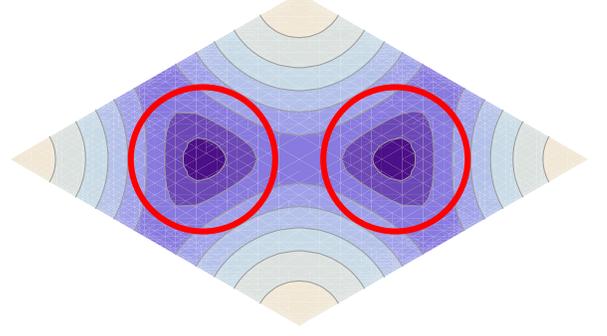}%
\end{center}
\caption{Contours of the tight binding energy $\protect\varepsilon _{\mathbf{%
k}}$ in the entire Brillouin zone of the honeycomb lattice. Circles of
radius $\mathcal{K}$ around the two Dirac points are the parts described by
the effctive low energy Weyl model. }
\end{figure}

The conductivity includes two apparently linearly divergent in time
"acceleration" parts. However their sum integrated over the BZ vanishes
since it is a full derivative of a periodic function, $Re\sum\nolimits_{%
\mathbf{k}}\left( h_{\mathbf{k}}h_{\mathbf{k}}^{\prime \prime }+\zeta _{%
\mathbf{k}}^{2}\right) /\varepsilon _{\mathbf{k}}=4\sum\nolimits_{\mathbf{k}%
}\varepsilon _{\mathbf{k}}^{\prime \prime }=0$. This cancellation, albeit,
is nontrivial: contributions come from the \textit{whole} BZ. When one uses
the effective Dirac theory, $h_{\mathbf{k}}=v_{0}\hbar \left( k_{x}\pm
ik_{y}\right) =v_{0}\hbar ke^{\pm i\varphi _{\mathbf{k}}},$ $\zeta _{\mathbf{%
k}}\approx v_{0}\hbar \cos \varphi _{\mathbf{k}}$ and integration over each
of the two circles (with the cutoff radius $\mathcal{K}$) in Fig.2 gives a
positive UV divergent result:

\begin{equation}
\int_{k=0}^{\mathcal{K}}k\int_{\varphi =0}^{2\pi }Re\left[ \frac{h_{\mathbf{k%
}}h_{\mathbf{k}}^{\prime \prime }+\zeta _{\mathbf{k}}^{2}}{\varepsilon _{%
\mathbf{k}}}\right] =v_{0}\hbar \int_{k=0}^{\mathcal{K}}\int_{\varphi
=0}^{2\pi }\cos ^{2}\varphi =\pi v_{0}\hbar \mathcal{K}\text{.}  \label{Int}
\end{equation}%
This is canceled exactly by contributions from regions of BZ far from Dirac
points in which the low energy effective model is not valid\cite{Kao}. Now
that the "acceleration" parts have cancelled, the oscillating term in $j_{p}$
at large $t$ limit can be safely calculated from the effective low energy
theory. Indeed averaging over long times $T=1/\eta $ by $C=\eta
\int_{t=0}^{\infty }C\left( t\right) e^{-\eta t}$, one obtains 
\begin{equation}
C^{0}=\frac{2v_{0}}{\pi ^{2}}\int_{k=0}^{\infty }\frac{\eta }{%
v_{0}^{2}k^{2}+\eta ^{2}}\int_{\varphi =0}^{2\pi }\cos ^{2}\varphi =1\text{,}
\label{sig}
\end{equation}%
as expected. Now we turn to the interaction corrections.

Similarly as before, the linear in $t$ "acceleration" corrections,

\begin{eqnarray}
C_{d} &=&-\frac{t}{\hbar S^{2}}\sum\limits_{\mathbf{p,q}}\frac{v_{\mathbf{p}-%
\mathbf{q}}}{\varepsilon _{\mathbf{q}}}Im\left( h_{\mathbf{q}}^{\prime
\prime }z_{\mathbf{q}}\right) Im\left( z_{\mathbf{q}}^{\ast }z_{\mathbf{p}%
}\right) ;  \label{sig1} \\
C_{p} &=&\frac{t}{\hbar S^{2}}\sum\limits_{\mathbf{p,q}}v_{\mathbf{p}-%
\mathbf{q}}\zeta _{\mathbf{q}}Re\left[ \left( \zeta _{\mathbf{q}}-\zeta _{%
\mathbf{p}}-i\frac{4\varepsilon _{\mathbf{q}}^{\prime }}{\varepsilon _{%
\mathbf{q}}}\right) \left( z_{\mathbf{q}}^{\ast }z_{\mathbf{p}}\right) %
\right] \text{,}  \notag
\end{eqnarray}
cancel each other beyond the Weyl model applicability domain, leaving
oscillating terms of $C_{p}$ that now take a form (averaged over large
times):%
\begin{eqnarray}
C &=&\frac{\eta }{\hbar S^{2}}\sum\limits_{\mathbf{p,q}}\frac{v_{\mathbf{p}-%
\mathbf{q}}\zeta _{\mathbf{q}}}{4\varepsilon _{\mathbf{q}}^{2}+\eta ^{2}}
\label{sig1final} \\
&&\left\{ 
\begin{array}{c}
\left[ \frac{\zeta _{\mathbf{q}}\left( 4\varepsilon _{p}^{2}+4\varepsilon _{%
\mathbf{q}}^{2}+\eta ^{2}\right) }{4\varepsilon _{\mathbf{p}}^{2}+\eta ^{2}}-%
\frac{\zeta _{\mathbf{p}}\left( 12\varepsilon _{\mathbf{q}}^{2}+\eta
^{2}\right) }{4\varepsilon _{\mathbf{q}}^{2}+\eta ^{2}}\right] Re\left( z_{%
\mathbf{q}}^{\ast }z_{\mathbf{p}}\right) \\ 
+\frac{2\varepsilon _{\mathbf{q}}^{\prime }}{\varepsilon _{\mathbf{q}}}%
Im\left( z_{\mathbf{q}}^{\ast }z_{\mathbf{p}}\right) +\frac{4\zeta _{\mathbf{%
p}}\varepsilon _{\mathbf{q}}\varepsilon _{\mathbf{p}}}{4\varepsilon _{%
\mathbf{p}}^{2}+\eta ^{2}}%
\end{array}%
\right\}  \notag \\
&=&\frac{11}{6}-\frac{\pi }{2}=C^{\left( 3\right) }.  \notag
\end{eqnarray}%
The integrals are again computed (see Supplemental material for details)
using the Dirac point approximation. This is the main result of the present
work. In this manner we also calculated the AC conductivity of the tight
binding model and results will be presented elsewhere.

\textit{Summary and discussion}. To summarize, we have calculated the
electron-electron interaction contribution to DC and AC conductivity of
undoped graphene within the tight binding model. Thus the controversy of
what is the actual magnitude (even order of magnitude) of the corrections is 
\textit{resolved} in favour of the intermediate value of the constant $%
C=C^{\left( 3\right) }>>C^{\left( 2\right) }$. It is shown that the
ambiguity between the three values originates in a nontrivial feature of
massless fermions, the chiral anomaly. The major complication that massless
fermions cause is\ the absence of a perfect scale separation between high
energies (on atomic scale $\gamma $) and low energies (effective Weyl theory
on the condensed matter scale $<<\gamma $). We demonstrated that some
aspects of the linear response physics are \textit{not} dominated by the two
Dirac points of the Brillouin zone at which the effective low energy model
is valid. For example, large contributions (infinite, when the size of the
Brillouin zone is being considered infinite) to the conductivity from the
vicinity of the Dirac points are cancelled by contributions from the region
between them. Another famous consequence of this scale nonseparation is the
"species doubling" of lattice fermions\cite{Smit}, which in the context of
graphene means that there necessarily appears a pair of Dirac points of
opposite chirality. The UV regularization of the effective theory \textit{%
does matter} and, if one were to use such a model, the only regularization
known to date to be consistent with the tight binding is the space
dimensional regularization developed in ref. \cite{Herbut2}. The reason is
not clear to us (especially due to the fact that fully relativistic
dimensional regularization in 2+1 anomalous theories is known to be
problematic\cite{Smit,anomaly}), but experiece with field theory would
indicate that one can also construct a successful sufficiently simple Pauli
- Villars kind of regularization.

If the result $C=C^{\left( 2\right) }$ were the correct one, the physics
would look very different. Indeed such a small value would easily explain
the experimenal absence of interaction corrections\cite{Schmalian2} in the
AC conductivity of graphene on a substrate\cite{GeimScience08}. The
explanation probably resides elsewhere, for example in the dielectric
constant of the substrate, screening due to puddles, etc\cite{Katsnelson2}.
We have calculated the effect of screened interactions represented by the
Hubbard model with quasilocal interactions (up to several nearest
neighbours) and obtained a vanishing first order correction to AC
conductivity at all frequencies in accordance with a general theorem\cite%
{Giuliani}. For local interactions this has been already noted in ref. \cite%
{MacDonald}.

The intermediate value of $C$ can also have a bearing on the putative
exciton condensation due to strong Coulomb interaction that has not been yet
experimentally observed even in suspended graphene samples\cite{Geim11} and
on interaction corrections to the dispersion relation of the excitations.
The random phase approximation (RPA) and various large $N_{f}$ results\cite%
{Castro3} should be also derived from the tight binding model or from a
properly regularized effective low energy one. It is well known in field
theory that generally chiral anomaly effects appear only in one loop
calculations \cite{anomaly} and higher orders resummed in RPA or $1/N_{f}$
approximation should not lead to further ambiguities. It is remarkable to
note that differences between the values of $C$ in equations Eqs.(\ref{c1},%
\ref{c2},\ref{c3}) are $C^{\left( 1,2\right) }=C^{\left( 3\right) }\pm 1/4$.
Sometimes due to anomalies similar differences in regularizations are
related to certain "topological invariant" stemming from the measure of the
path integration over fermionic fields\cite{anomaly}. Here the situation is
more complicated since we are dealing with correction due to interactions,
not with the simple bubble diagram.

Acknowledgements. We are indebted to Y. Yaish, H.C. Kao, E. Andrei, V.
Nazarov and W.B. Jian for valuable discussions.


\begin{thebibliography}{99}
\bibitem{remark} Here we have restricted ourselves to calculate the
correction to the leading order in $\alpha $. Of course, beyond this order
the effects of Fermi velocity renormalization appear. When performing the
renormalization group improvement of the correction one takes the value of
the Fermi velocity at the appropriate scale.
\end{thebibliography}
\end{document}